\documentclass[12pt,preprint]{aastex}
\usepackage{rotating}
\usepackage{wasysym}
\shorttitle{Spitzer confirms the transit of HD\,97658\,b}
\shortauthors{V. Van Grootel et al.}

\begin{document}
\title{Transit confirmation and improved stellar and planet parameters for the super-Earth HD\,97658\,b and its host star}%\thanks{The photometric time series used in this work are only available in electronic form at the CDS via anonymous ftp to cdsarc.u-strasbg.fr (130.79.128.5) or via http://cdsweb.u-strasbg.fr/cgi-bin/qcat?J/A+A/}}
\author{V. Van Grootel\altaffilmark{1}\footnote{Charg\'ee de recherches, Fonds de la Recherche Scientifique, FNRS, rue d'Egmont 5, B-1000 Bruxelles, Belgium.}, M. Gillon\altaffilmark{1}, D. Valencia\altaffilmark{2}, N.~Madhusudhan\altaffilmark{3}, D. Dragomir\altaffilmark{4,5}, A.\,R. Howe\altaffilmark{6}, A.\,S. Burrows\altaffilmark{6}, B.-O.~Demory\altaffilmark{7,3}, D.~Deming\altaffilmark{8}, D. Ehrenreich\altaffilmark{9}, C.~Lovis\altaffilmark{9}, M.~Mayor\altaffilmark{9}, F.~Pepe\altaffilmark{9}, D.~Queloz\altaffilmark{3,9}, R. Scuflaire\altaffilmark{1}, S.~Seager\altaffilmark{7}, D.~Segransan\altaffilmark{9}, S.~Udry\altaffilmark{9}}

\altaffiltext{1}{Institut d'Astrophysique et de G\' eophysique, Universit\' e
  de Li\` ege, 17 All\'ee du 6 Ao\^ ut, B-4000 Li\` ege, Belgium;valerie.vangrootel@ulg.ac.be}
 \altaffiltext{2}{Department of Physical and Environmental Sciences, University of Toronto, 1265 Military Trail, Toronto, ON, M1C 1A4, Canada}
\altaffiltext{3}{Institute of Astronomy, University of Cambridge, Madingley Road, Cambridge, CB3 0HA, UK}
\altaffiltext{4}{Las Cumbres Observatory Global Telescope Network, 6740 Cortona Dr. suite 102, Goleta, CA 93117, USA}
\altaffiltext{5}{Department of Physics, Broida Hall, UC Santa Barbara, CA, USA}
\altaffiltext{6}{Department of Astrophysical Sciences, Princeton University, Princeton, NJ 08544, USA}
\altaffiltext{7}{Department of Earth, Atmospheric and Planetary Sciences, Department of Physics, Massachusetts Institute of Technology, 
77 Massachusetts Ave., Cambridge, MA 02139, USA}
\altaffiltext{8}{Department of Astronomy, University of Maryland, College Park, MD 20742-2421, USA}
\altaffiltext{9}{Observatoire de Gen\`eve, Universit\'e de Gen\`eve, 51 Chemin des Maillettes, 1290 Sauverny, Switzerland}

\newpage

\begin{abstract}Super-Earths transiting nearby bright stars are key objects that simultaneously allow for accurate measurements of both their mass and radius, providing essential constraints on their internal composition. We present here the confirmation, based on \textit{Spitzer} transit observations, that the super-Earth HD\,97658\,b transits its host star. HD\,97658 is a low-mass ($M_*=0.77\pm0.05\,M_{\odot}$) K1 dwarf, as determined from the \textit{Hipparcos} parallax and stellar evolution modeling. To constrain the planet parameters, we carry out Bayesian global analyses of Keck-HIRES radial velocities, and \textit{MOST} and \textit{Spitzer} photometry.  HD\,97658\,b is a massive ($M_P=7.55^{+0.83}_{-0.79} M_{\oplus}$) and large ($R_{P} = 2.247^{+0.098}_{-0.095} R_{\oplus}$ at 4.5 $\mu$m) super-Earth. We investigate the possible internal compositions for HD\,97658\,b. Our results indicate a large rocky component, by at least 60\% by mass, and very little H-He components, at most 2\% by mass. We also discuss how future asteroseismic observations can improve the knowledge of the HD\,97658 system, in particular by constraining its age. Orbiting a bright host star, HD\,97658\,b will be a key target for coming space missions TESS, CHEOPS, PLATO, and also JWST, to characterize thoroughly its structure and atmosphere. 
\end{abstract}  


\section{Introduction}
Among the diversity of the exoplanetary population, the class of the so-called ``super-Earths", with a mass of a few Earth masses, is of utmost interest. Firstly, they do not exist in the Solar System, leaving us with only extrasolar planets for detailed studies. Secondly, they seem to be extremely common in the Galaxy \citep{2013ApJ...766...81F,2013ApJ...770...69P,2013Sci...340..572H}. Thirdly, a wide variety of interiors are theoretically possible for a small range of radii ($1 < R_{\rm P}/R_{\oplus} < 5$) and masses ($1 < M_{\rm P}/M_{\oplus} < 10$), from pure iron composition with no atmosphere, to water planets with a significant atmosphere \citep{2007ApJ...665.1413V,2007ApJ...669.1279S}. Transiting super-Earths around bright nearby stars are therefore key objects that allow for accurate measurements of both their mass and radius \citep{2012A&A...539A..28G}. Equally important, they also allow for atmospheric measurements from transit transmission spectroscopy (e.g. \citealt{2012ApJ...747...35B}) and/or occultation (secondary eclipse) emission spectroscopy \citep{2012ApJ...751L..28D} without the challenging task of resolving the planet's light from the one of its host star. Such direct imaging is only possible to date for widely separated, massive young planets \citep{2010Natur.468.1080M,2013A&A...559L..12A,2013ApJ...774...11K}. Transiting super-Earths around bright nearby stars are therefore essential to shed light on their true nature and origin.

HD\,97658 is the second brightest ($V=7.7$, $K=5.7$) host star found to be transited by a super-Earth (the brightest being 55 Cnc; \citealt{2011ApJ...737L..18W,2011A&A...533A.114D}). HD\,97658\,b was detected from radial velocity (RV) measurements with Keck-HIRES spectroscopy, with a minimum mass $M_{\rm P} \sin i = $ 8.2 $\pm$ 1.2 $M_{\oplus}$ and an orbital period $P=9.494 \pm 0.005$ days \citep{2011ApJ...730...10H}. The detection of transits from ground-based observations was announced \citep{2011arXiv1109.2549H}, but follow-up space-based \textit{MOST} photometry showed no transits at the claimed ephemeris \citep{2012ApJ...759L..41D}. Further \textit{MOST} observations finally discovered transit events with an ephemeris consistent with that predicted from RV measurements \citep{2013ApJ...772L...2D}. From a global Markov Chain Monte-Carlo (MCMC) analysis simultaneously modeling HIRES RVs and \textit{MOST} photometry, they derived a 2.34$^{+0.18}_{-0.15}\,R_{\oplus}$ and 7.86 $\pm$ 0.73 $M_{\oplus}$ planet, confirming its super-Earth nature with an average density of 3.44$^{+0.91}_{-0.82}$ g cm$^{-3}$.

The announcement of transit detections by \citet{2013ApJ...772L...2D} motivated us to include HD\,97658\,b in our \textit{Spitzer} program to search for and observe transits of low-mass RV-detected planets \citep{2011IAUS..276..167G}. HD\,97658\,b was observed in August 2013 with the \textit{Spitzer} Infra-Red Array Camera (IRAC, \citealt{2004ApJS..154...10F}) at 4.5 $\mu$m. \textit{Spitzer} already proved to be the tool of choice to perform exquisite precise and continuous photometry for the detection and characterization of the ultra-shallow transits of super-Earths, such as HD\,40307\,b (for which transits are firmly discarded; \citealt{2010A&A...518A..25G}) and 55 Cnc e (\citealt{2011A&A...533A.114D}). We present here the confirmation of the transiting nature of HD\,97658\,b from \textit{Spitzer} observations. We also present global Bayesian analyses of Keck-HIRES RV data, and \textit{MOST} and \textit{Spitzer} photometry, in order to derive as accurately as possible the parameters of HD\,97658\,b. 

Ultimately, the precision for planetary parameters is related to the precision achieved for the stellar parameters, since planetary masses and radii as determined by transit and RV measurements cannot be determined independently from the properties of their host stars. Atmospheric stellar parameters such as effective temperature $T_{\rm eff}$ and metallicity are determined from spectroscopy. The stellar radius, at least for nearby stars, can be derived from the luminosity (from parallax measurements) and the spectroscopic effective temperature using the classical equation of stellar physics $L_*=4\pi R_*^2 \sigma T_{\rm eff}^4$. Obtaining the stellar mass is a more delicate process that can be achieved by a stellar evolution code using $T_{\rm eff}$, metallicity and luminosity as inputs. The stellar age is usually poorly constrained in this process \citep{2010ARA&A..48..581S}. A powerful method to provide accurate and precise stellar radius, mass and age is asteroseismology, by modeling the oscillation spectrum of the star. The CoRoT \citep{2006ESASP1306...33B} and \textit{Kepler} \citep{2010ApJ...713L..79K} space missions were based on this complementarity, while the future TESS \citep{2010AAS...21545006R} and PLATO \citep{2013Rauer} space missions will also include significant asteroseismic programs. We also investigate here how future asteroseismic observations of HD\,97658 can help to further improve the knowledge of the star and planet parameters. 

This paper is organized as follows. The \textit{Spitzer} observations and their reduction are presented in Sect.~\ref{data}, while host star modeling is presented in Sect.~\ref{host}. Section~\ref{Spitzer} is devoted to the analysis of the \textit{Spitzer} data, while Sect.~\ref{Analysis} presents global Bayesian analyses of RV, \textit{MOST} and \textit{Spitzer} data. We discuss the results and provide avenues to further improve the characterization of HD\,97658\,b and its host star in Sect.~\ref{discuss}. Conclusion and prospects are given in Sect.~\ref{cc}.

\section{\textit{Spitzer} photometry}
\label{data}
We  monitored HD\,97658 with {\it Spitzer}'s IRAC camera on 2013 Aug 10 from 13h01 to 18h27 UT, corresponding to a transit window\footnote{the "transit window" is the window of time within which the transit is likely to occur.} as computed from the MOST transit ephemeris \citep{2013ApJ...772L...2D}. These {\it Spitzer} data were acquired in the context of the Cycle 9 program 90072 (PI: M. Gillon) dedicated to the search for the transits 
of  RV-detected low-mass planets. They consist of 2320 sets of 64 individual subarray images obtained at 4.5 $\mu$m with an integration time of 0.08 s. They are available on the  {\it Spitzer} Heritage Archive database\footnote{http://sha.ipac.caltech.edu/applications/Spitzer/SHA} under the form of 2320 Basic Calibrated Data (BCD) files calibrated by the standard {\it Spitzer} reduction pipeline (version S19.1.0). 

\begin{figure}[!h]
\begin{center}
 \includegraphics[scale=0.6,angle=0]{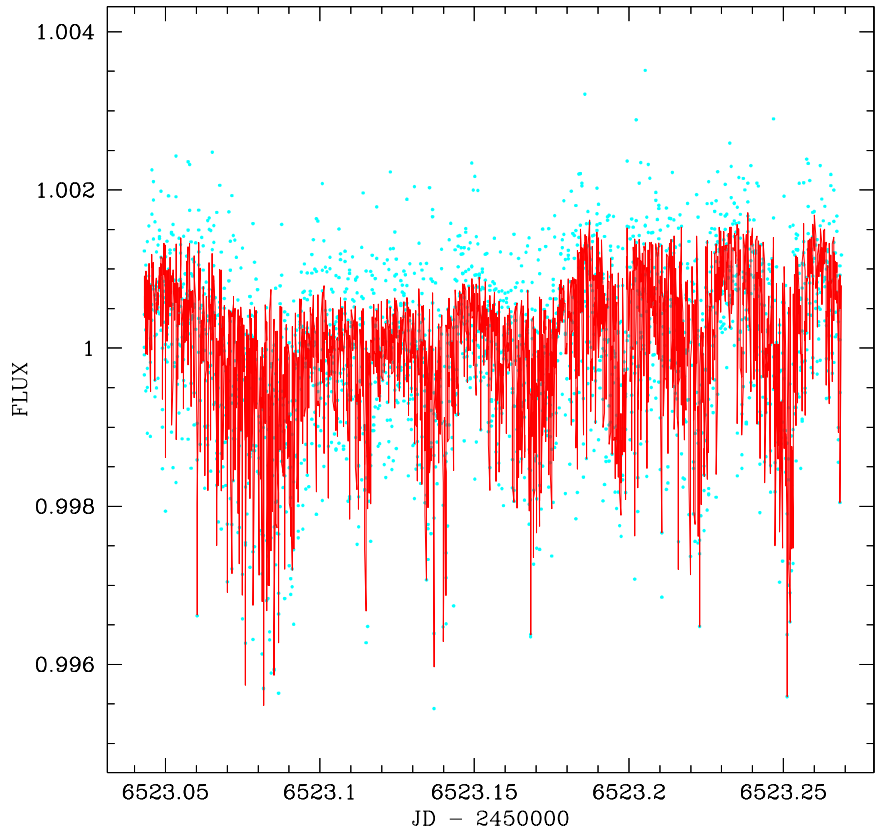} 
 \caption{Raw HD\,97658 \textit{Spitzer} IRAC light curve (blue dots), with the global photometric model overimposed (red curve) made of the photometric eclipse model of  \citet{2002ApJ...580L.171M} multiplied by the baseline model representing the \textit{Spitzer} instrumental effects. See text for details.}
   \label{raw}
\end{center}
\end{figure}

We first converted fluxes from the {\it Spitzer} units of specific intensity (MJy/sr) to photon counts, then aperture photometry was performed on each subarray image with the {\tt IRAF/DAOPHOT}\footnote{IRAF is distributed by the National Optical Astronomy Observatory, which is operated by the Association of Universities for Research in Astronomy, Inc., under cooperative agreement with the National Science Foundation.} software \citep{1987PASP...99..191S}. We tested different aperture radii and background annuli, the best result (the one that gives the lowest white noise, by minimizing the rms of the residuals, and the lowest red noise, by following the approach presented in \citealt{2006A&A...459..249G}) being obtained with an aperture radius of 3 pixels and a background annulus extending from 11 to 15.5 pixels from the point-spread function (PSF) center. The PSF center was measured by fitting a 2D-Gaussian profile on each image. The $x$-$y$ distribution of the measurements was examined, and we discarded the few measurements having a very different position than the bulk of the data. For each block of 64 subarray images, we discarded the discrepant values for the measurements of flux, background, $x$ and $y$ positions using a $\sigma$ median clipping (5-$\sigma$ for the flux and 10-$\sigma$ for the other parameters), and the resulting values were averaged, the photometric error being taken as the error on the average flux measurement. A 20-$\sigma$ running median clipping was used on the resulting light curve to discard totally discrepant fluxes (due, e.g., to cosmic rays). In the end, only 0.05\% of the measurements were rejected. The blue dots of  Fig.~\ref{raw} represent the resulting IRAC photometric light curve of HD\,97658. These data, which are directly used in our MCMC algorithm (see Sect.~4 and 5), can be found on Tab.~\ref{rawdata} (the full version is available online in the form of a machine-ready table).

\section{Host star parameters}
\label{host}
An accurate knowledge of the host star, including the stellar mass and radius, is needed in any exoplanetary modeling. In addition, the age of the star is an excellent proxy for the age of its planets, since they are expected to have formed within a few million years of each other. Two sources are available for the atmospheric stellar parameters of HD\,97658 (\citealt{2011ApJ...730...10H} and \citealt{2011arXiv1109.2549H}). From Howard et al. 2011 (resp. \citealt{2011arXiv1109.2549H}), the effective temperature is $T_{\rm eff} = 5170 \pm 44$ K (resp. $T_{\rm eff} = 5119 \pm 44$ K), the surface gravity log $g =$ 4.63 $\pm$ 0.06 (resp. 4.52 $\pm$ 0.06), and [Fe/H] $= -0.23 \pm 0.03$ (resp. $-0.30 \pm 0.03$). We assumed here that [Fe/H] represents the global metallicity with respect to the Sun, defined as $[\log (Z/X)_* - \log (Z/X)_{\odot}]$, where $X$ and $Z$ are the fractional mass of hydrogen and elements heavier than helium respectively. A \textit{Hipparcos} parallax measurement of HD\,97658 is available: $\pi=$ 47.36 $\pm$ 0.75 mas \citep{2007A&A...474..653V}, giving a distance of 21.11 $\pm$ 0.34 pc. Using observed magnitudes of \citet{2010MNRAS.403.1949K} and bolometric corrections of \cite{1996ApJ...469..355F} (with $M_{\rm bol,\odot} = 4.73$, \citealt{2010AJ....140.1158T}), this translates to a stellar luminosity of $L_*/L_{\odot} = 0.355 \pm 0.018$. As is standard, all the errors cited in this section and throughout the paper are the 1-$\sigma$ range (68.3\% probability).

We used the effective temperature, metallicity and luminosity with their respective errors as inputs for stellar evolution modeling with the CLES (Code Li\'egeois d'Evolution Stellaire) code \citep{2008Ap&SS.316...83S}. In all evolutionary computations we used the mixing-length theory (MLT) of convection \citep{1958ZA.....46..108B} and the CEFF equation of state \citep{1992A&ARv...4..267C}. We considered here the most recent solar mixture\footnote{"mixture" is stellar physics jargon that means the proportions of $X$, $Y$, and $Z$, of course with $X+Y+Z=1$.} of \citet{2009ARA&A..47..481A}, giving for the present Sun $(Z/X)_{\odot} = 0.0181$. We used, for this metallic mixture, the OPAL opacity tables \citep{1996ApJ...464..943I} for the high temperatures and the ones of \citet{2005ApJ...623..585F} for the low temperatures. The surface boundary conditions are given by atmospheres computed within the Eddington approximation. Microscopic diffusion (gravitational settling) is taken into account, but no radiative accelerations of metals were included given the low mass we expect for the host star \citep{2012A&A...543A..96E}. For the same reason, no convective core is expected, so no overshooting was considered here. Finally, the $\alpha$ parameter of the MLT was kept fixed to the solar calibration ($\alpha_{\rm MLT} = 1.8$). Since the helium atmospheric abundance cannot be directly measured from spectroscopy in low-mass stars such as HD\,97658, we computed evolutionary tracks with three initial helium abundances: the solar value, %($Y_{\odot}=0.2485$), 
a value labelled $Y_G$ that increases with $Z$ (as expected if the local medium follows the general trend observed for the chemical evolution of galaxies; \citealt{2010ApJ...710L..67I}), and an arbitrary %higher 
value $Y_{\rm arb}=0.26$ close to the most recent primordial value from the Big Bang nucleosynthesis \citep{2010ApJ...710L..67I}.

\begin{figure}[!h]
\begin{center}
 \includegraphics[scale=0.6,angle=0]{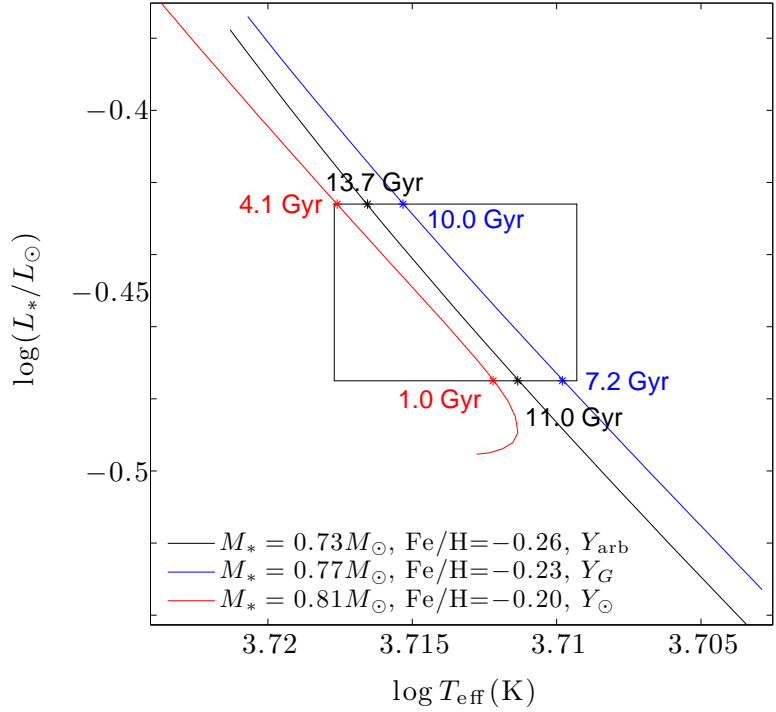} 
 \caption{Evolutionary tracks in a $\log T_{\rm eff} - \log (L_*/L_{\odot}$) diagram for HD 97658, for various masses and metallicities that respect the observational constraints ($L_*$, $T_{\rm eff}$, [Fe/H]). Several initial mixtures, in particular for the helium abundance ($Y_{\odot}$, $Y_G$, and $Y_{\rm arb}$) were also considered (see text for details). The age of the star when it crosses the 1-$\sigma$ box $\log T_{\rm eff} - \log (L_*/L_{\odot}$) is also indicated.}
   \label{hr}
\end{center}
\end{figure}

We computed many evolutionary tracks for several masses with the constraints of having consistent metallicities and effective temperatures as given by spectroscopy, and having luminosities consistent with the one derived from the \textit{Hipparcos} parallax. No stellar models younger than the Universe were found to have a luminosity $L_*/L_{\odot} = 0.355 \pm 0.018$ within the 1-$\sigma$ range of effective temperature and metallicity given by \citet{2011arXiv1109.2549H}, which are therefore discarded from stellar evolution modeling. Within the 1-$\sigma$ range given by \citet{2011ApJ...730...10H} for the metallicity and effective temperature, the values of stellar mass are $M_* = 0.77 \pm 0.05$ $M_{\odot}$ (Fig.~\ref{hr}). Unfortunately, no useful constraints on the stellar age were obtained: such stars can reach $L_*/L_{\odot} = 0.355 \pm 0.018$ when they are a few Gyr old only if they are $\sim$ 0.82 $M_{\odot}$, but they can be much older, up to the age of the Universe, if they are less massive (see Fig.~\ref{hr}). This also strongly depends on the initial mixture. All evolutionary tracks that respect the observational constraints on $T_{\rm eff}$, $L_*$ and [Fe/H] correspond to stars that are on the main sequence, in the H-core burning phase.

Combining the stellar luminosity $L_*/L_{\odot} = 0.355 \pm 0.018$ with the effective temperature of \citet{2011ApJ...730...10H} results in a stellar radius $R_*/R_{\odot} = 0.74 \pm 0.03$. These values are somewhat different from the ones cited by \citet{2011ApJ...730...10H}, and also by \citealt{2011arXiv1109.2549H} (which were taken by \citealt{2013ApJ...772L...2D}). The discrepancy comes from their announced luminosity $L_*/L_{\odot} = 0.30 \pm 0.02$ and absolute magnitude $M_V = 6.27 \pm 0.10$, which are not consistent with the \textit{Hipparcos} parallax (\citealt{2007A&A...474..653V}; see also Table 1 of \citealt{2010MNRAS.403.1949K}, HD\,97658 $\equiv$ HIP 54906). 

\section{\textit{Spitzer} confirms the transiting nature of HD\,97658\,b}
\label{Spitzer}
The \textit{Spitzer} light curve was analyzed with the adaptative MCMC algorithm presented in \citet[][and references therein]{2012A&A...542A...4G}, with the aim to confirm/refute the transiting nature of HD\,97658\,b and the ephemeris of \citet{2013ApJ...772L...2D}. MCMC is a Bayesian inference method based on stochastic simulations that sample the posterior probability distributions of adjusted parameters for a given model. Our MCMC implementation uses the Metropolis-Hasting algorithm (see, e.g., \citealt{2008Baye.book}) to perform this sampling. Our nominal model was based on a star and a transiting planet on a Keplerian orbit about their center of mass. We modeled the eclipse photometry with the photometric eclipse model of \citet{2002ApJ...580L.171M}, multiplied by a baseline model representing the low-frequency instrumental effects. This baseline model was a sum of a second-order polynomial in the $x$- and $y$-positions of the PSF center, first-order polynomials in the width of the PSF in the $x$- and $y$-direction, and a first-order polynomial of the logarithm of time. The two first kinds of polynomial aimed to model the ``pixel-phase effect'' affecting the IRAC InSb arrays, while the last one  represented the sharp increase of the {\it Spitzer} detectors' response, generally called ``the ramp'' in the literature (see, e.g.,  \citealt{2008ApJ...673..526K}). This baseline model was elected based on the minimization of the Bayesian Information Criterion (BIC; \citealt{Schwarz}). The final global photometric model (eclipse model multiplied by the baseline model) is shown by the red curve of  Fig.~\ref{raw}, which nicely illustrates the importance of detrending operations on \textit{Spitzer} raw photometry in order to extract the scientific information. Finally, quadratic limb-darkening was assumed, with values deduced from the tables of \citet{2011A&A...529A..75C} for the appropriate \textit{Spitzer} filters and stellar parameters. 

The jump parameters, i.e. the parameters over which the MCMC random walk (here made of 10 chains of $10^4$ steps) occurs, were: the transit depth $dF$, the transit width (from first to last contact) $W$, the transit timing (time of minimum light) $T_0$, and the impact parameter $b'=a \cos i / R_*$. We assumed a uniform prior distribution for all these jump parameters. The orbital period $P$ was also a jump parameter, but this time with a Gaussian prior based on the \citet{2013ApJ...772L...2D}'s value. In our MCMC implementation, it means that we imposed a Bayesian penalty:
\begin{equation}
\label{eqn1}
BP_{ephemeris}=\frac{(P-P_{D13})^2}{\sigma^2_{P_{D13}}}
\end{equation}
where $P_{D13} =9.4909$ d and $\sigma_{P_{D13}} = 0.0016$ d \citep{2013ApJ...772L...2D}. Gaussian priors were also imposed on the stellar mass, luminosity, effective temperature, and metallicity (see Sect.~\ref{host}), i.e. we also applied a Bayesian penalty similarly to eq.~(\ref{eqn1}) for these parameters. The merit function used in our MCMC simulation was therefore the sum of the $\chi^2$ for each data set (quadratic sum of the difference between the model and the data) and of the Bayesian penalties. It is also important to include stellar parameters with their Gaussian priors in the MCMC simulation to properly propagate the errors and to accurately derive the physical parameters from the jump parameters. If a jump parameter is not constrained by observational data, its a posteriori distribution will be the same as its a priori distribution, but we can use the resulting distribution to propagate errors correctly.

\begin{figure}[!h]
\begin{center}
 \includegraphics[scale=0.6,angle=0]{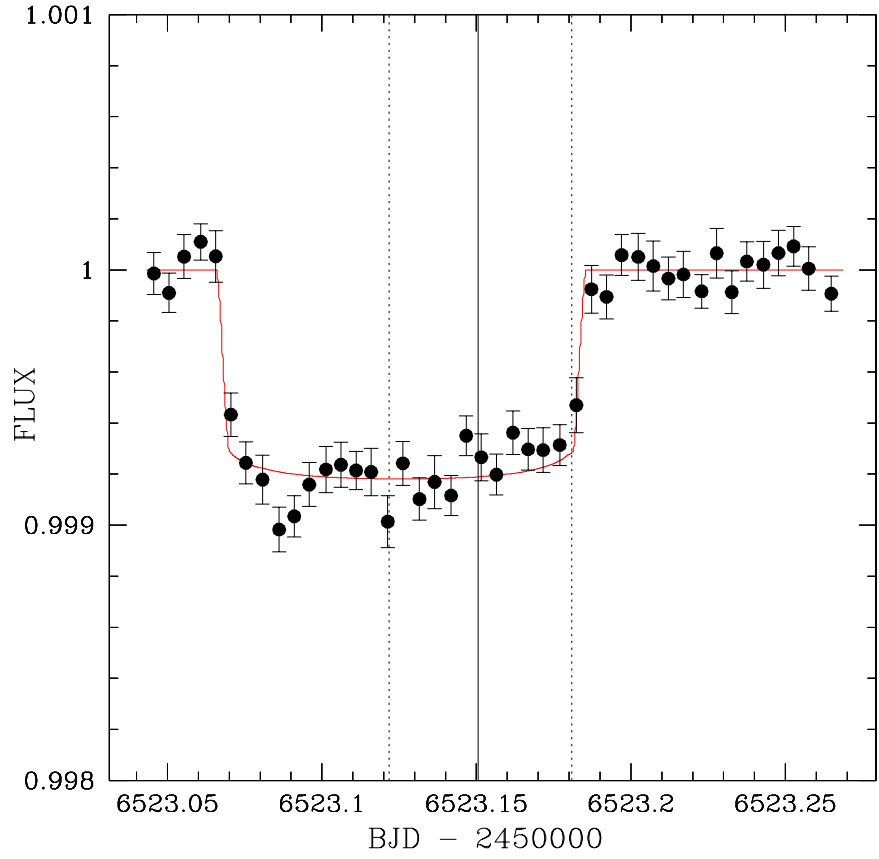} 
 \caption{HD\,97658 \textit{Spitzer} IRAC light curve binned to intervals of five minutes, with the best-fit transit model overimposed. The vertical solid line is the propagated mid-transit time of \citet{2013ApJ...772L...2D}, with its 1-$\sigma$ errors (dashed vertical lines).}
   \label{fig1}
\end{center}
\end{figure}

The good convergence and the quality of the sampling of the MCMC simulation were successfully checked using the statistical test of \citet{GelmanRubin}, by verifying that the so-called potential reduction factors for all jump parameters are close to unity (within 1\%). Figure~\ref{fig1} shows the \textit{Spitzer} IRAC photometry corrected for the systematics and binned into five-minute-wide intervals with the best-fitting eclipse model superimposed. The vertical lines show the propagated (17 planetary orbits later) \textit{MOST} mid-transit time, with 1-$\sigma$ errors, of \citet{2013ApJ...772L...2D}. The \textit{Spitzer} detected transit at $T_0 = 2456523.12544^{+0.00062}_{-0.00059}$ (BJD\_TDB) thus fully confirms, within 1-$\sigma$, the ephemeris provided by \citet{2013ApJ...772L...2D}. 

\section{Global Bayesian analyses}
\label{Analysis}
In order to get the strongest constraints on the system parameters, we performed several global Bayesian analyses using as input data not only our {\it Spitzer} transit photometry, but also the detrended \textit{MOST} transit light curves \citep{2013ApJ...772L...2D} and the 171 published Keck-HIRES RVs acquired and reduced using the same techniques as in \citet{2011ApJ...730...10H} and compiled by \citet{2012ApJ...759L..41D,2013ApJ...772L...2D}.
\subsection{Keck-HIRES RV and \textit{Spitzer} photometry data}
\label{RVS}
We performed a MCMC simulation made of 10 chains of $5 \times 10^4$ steps including HIRES RV data and \textit{Spitzer} photometry. We used a classical Keplerian model for the RVs (no Rossiter-McLaughlin effect was detected). The jump parameters in our MCMC simulation were: the transit depth $dF$, the transit width (from first to last contact) $W$, the transit timing (time of minimum light) $T_0$, the impact parameter $b'=a \cos i / R_*$, the orbital period $P$, the two parameters $\sqrt{e} \cos \omega$ and $\sqrt{e} \sin \omega$ ($e$ is the eccentricity and $\omega$ the argument of periastron) and the $K_2$ parameter (related to the RV orbital semi-amplitude $K$ via $K_2 = K \sqrt{1-e^2}\,P^{1/3}$). We also allowed the quadratic limb-darkening coefficients $u_1$ and $u_2$ to float in this MCMC simulation, using as jump parameters not these coefficients themselves but the combinations $c_1 = 2 \times u_1 + u_2$ and $c_2 =  u_1 - 2 \times u_2$ to minimize the correlation of the obtained uncertainties \citep{2006ApJ...652.1715H}. The prior distributions of these limb darkening coefficients directly depend on the ones on the effective temperature and metallicity, for which Gaussian priors, i.e. Bayesian penalties similarly to eq.~(\ref{eqn1}), were imposed ($T_{\rm eff} = 5170 \pm 50$ K and [Fe/H] $= -0.23 \pm 0.03$). Gaussian priors were also imposed on the stellar mass ($M_*= 0.77 \pm 0.05 M_{\odot}$) and luminosity ($L_* = 0.355 \pm 0.018 L_{\odot}$). Again, it is important to include these parameters with their Gaussian priors in the MCMC simulation in order to correctly propagate the errors and to accurately derive the physical parameters from the jump parameters. 

Table \ref{tab1} shows the median values and 68.3\% probability interval for the jump parameters mentioned above given by our MCMC simulation of \textit{Spitzer} photometry and Keck-HIRES RVs, as well as the derived stellar and planetary physical parameters. The good convergence and the quality of the sampling of the 10 chains of the MCMC simulation were successfully checked using here again the Gelman \& Rubin statistical test (all jump parameters have potential reduction factors near unity within 1\%). 

\subsection{Keck-HIRES RV, \textit{MOST} and \textit{Spitzer} data}
Finally, we repeated the global Bayesian analysis, but this time including the three continuous MOST transits. We only used these three light curves acquired when the star was fully in the Continuous Viewing Zone of the satellite, the two other light curves showing a much poorer quality due to gaps in coverage and higher levels of correlated noise. As discussed in \citet{2013ApJ...772L...2D}, the \textit{MOST} reduction pipeline suppresses the depth of a transit signal, by an amount of $\sim$10\%, as indicated by transit injection tests in these 3 \textit{MOST} transits. To include this effect and the uncertainty on it, we assumed a dilution of $10\pm2$ \% for the \textit{MOST} photometry. We carried out a global MCMC analysis with the same number of steps and chains, the same jump parameters, the same Gaussian priors, and the same check of the convergence and sampling quality as in Sect.~\ref{RVS}. The most interesting result is the transit depth in visible, from \textit{MOST} photometry: $dF_{MOST}=949^{+81}_{-75}$ ppm. This translates to a planet radius $R_P = 2.49^{+0.14}_{-0.13} R_{\oplus}$, larger at the 2-$\sigma$ level than the planet radius at 4.5 $\mu$m derived from \textit{Spitzer} photometry (Tab.~\ref{tab1}). This may be related to some instrumental \textit{MOST} systematics not fully corrected, beyond the 2\% uncertainty we took here on the dilution effect. This could also be related to the atmospheric composition of HD\,97658\,b, as we discuss in the next section. We note that our planet radius based on \textit{MOST} photometry is slightly higher than the one found by \citealt{2013ApJ...772L...2D} (2.34$^{+0.18}_{-0.15}\,R_{\oplus}$), although within their 1-$\sigma$ uncertainties. This is directly related to the more luminous and, therefore, larger radius star we modeled (see Sect.~3).

\section{Discussion}
\label{discuss}
\subsection{HD\,97658\,b, a key object for super-Earth characterization}
\subsubsection{Internal composition of HD\,97658\,b}
Our MCMC results (see Table \ref{tab1}) give a planetary radius of $R_{P} = 2.247^{+0.098}_{-0.095}\,R_{\oplus}$ as measured in IRAC 4.5 $\mu$m channel, and a planetary mass of $M_P=7.55^{+0.83}_{-0.79}\,M_{\oplus}$.  This yields a super-Earth with an intermediate average density ($\rho_P=3.90^{+0.70}_{-0.61}$ g cm$^{-3}$), close to the average density of Mars ($\rho_{\mars}=3.9335$ g cm$^{-3}$). \citet{2006Icar..181..545V} proposed a model for the mass-radius relationship for rocky planets as $M_P = a R_P ^4$, where $a$ depends on the composition. For the same composition, the average density of a planet therefore increases as $\rho_P \sim M_P/R_P^3 \sim M_P^{0.25}$. Given that HD\,97658\,b is 75 times more massive than Mars, it clearly has a much lighter composition.

\begin{figure}[!h]
\begin{center}
 \includegraphics[scale=0.8,angle=0]{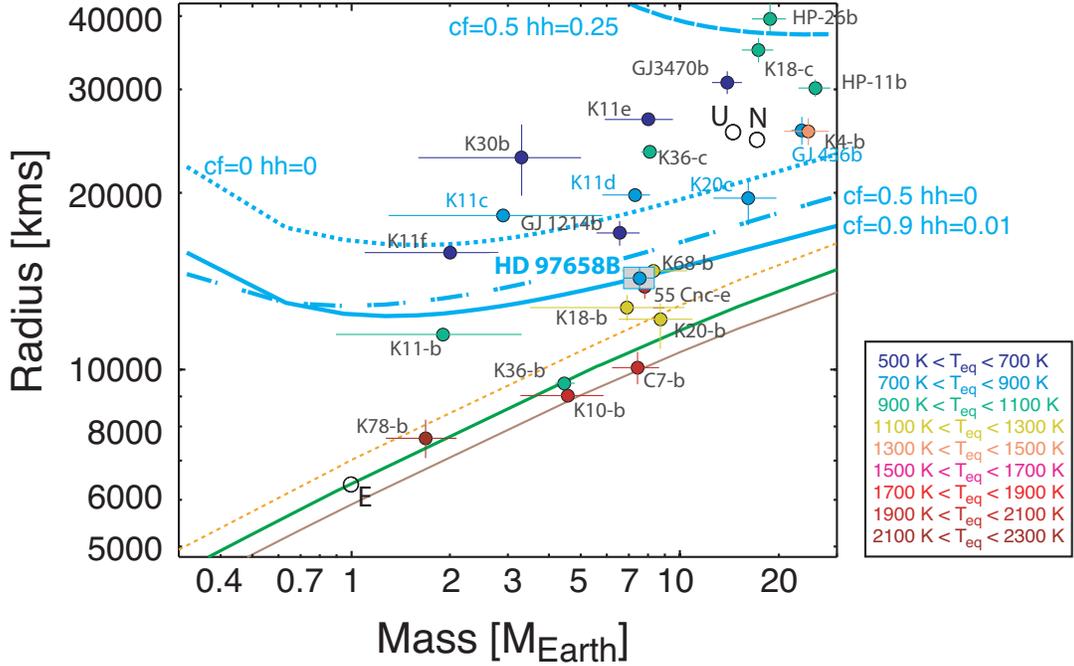} 
 \caption{Planetary mass-radius diagram of HD 97658\,b and other low-mass exoplanets.  The planets are color coded according to their equilibrium temperature.  The orange line is the MR relationship for a pure silicate planet (the lightest rocky composition), the green is an Earth-like composition (2/3 silicate mantle with 10\% iron by mol, 1/3 iron core), and the brown line is an iron-rich composition (37\% silicate mantle, 63\% iron core). The blue lines show specific MR relations. "Cf" stands for core fraction, while "hh" is for bulk hydrogen and helium by mass: (1) core fraction=0, H-He=0, and water/ices=1; (2) core fraction =0.5, H-He=0 and water/ices=0.5; (3) core fraction =0.5, H-He=0.25 and water/ices=0.25; (4) core fraction = 0.9, H-He=0.01 and water/ices=0.09.
 }
   \label{mr}
\end{center}
\end{figure}

In Fig.~\ref{mr} we show how HD\,97658\,b compares to the other detected low-mass planets for which we have measured mass and radius. It is worth noting that the mass and radius of HD\,97658\,b are very similar to Kepler 68 b \citep{2013ApJ...766...40G}. To illustrate that HD\,97658\,b has both a low H-He content and a high core-mass fraction, within the limitations of a mass-radius diagram, we show mass-radius relations (blue lines) for four specific compositions, calculated at $T_{\rm eq}=700$ K:
\begin{itemize}
\item (1): a water planet (no core and no H-He -- dotted blue line);
\item (2): core fraction of 50\% and no H-He (dashed dotted line);
\item (3): core-fraction of 50\% and H-He 25\% (long dashed line);
\item (4): core-fraction of 90\% and H-He = 1\% (solid line).
\end{itemize}

A water-only composition (1) sits above the data for HD\,97658\,b, even though it has no H-He.  If we consider larger core-mass fractions, only those above 50\% (2--4) start to approach the data for HD\,97658\,b. If we include any amount of H-He the amount of core needs to be larger. One possible composition for HD\,97658\,b is a planet with an earth-like core of 90\%, 1\% H-He, and 9\% water/ices (Fig.~\ref{mr}).

\begin{figure}[!h]
\begin{center}
 \includegraphics[scale=0.6,angle=0]{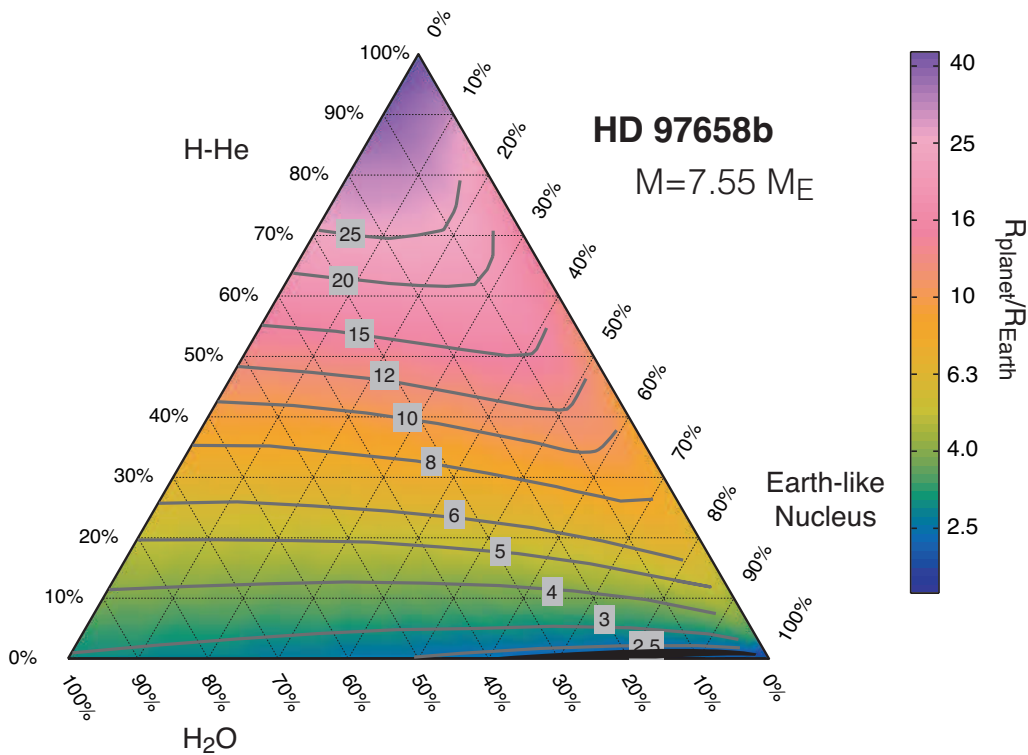} 
 \caption{Ternary Diagram for HD\,97658\,b. This diagram shows the radius for all possible compositions for a planet of a specific mass (in this case $7.55\, M_{\oplus})$.  Each point depicts a unique planetary composition from a combination of H-He, H$_2$O + ices, and rocky earth-like nucleus (33\% iron core below a 67\% silicate mantle.  Each vertex corresponds to a 100\% of each compositional end member (with H-He on top, H$_2$O + ices in the low left corner and earth-like nucleus on the low right corner), and 0\% on the opposite line.  The color bar shows the radius in terms of Earth radii, and the gray lines are the isoradius curves labeled in terms of Earth radii. The black shaded region in the low right corner shows the possible compositions for HD\,97658\,b and the width of the shaded region takes into account the uncertainty in the mass and radius (for all combinations between $M+\Delta M$, $R-\Delta R$ and $M-\Delta M$, $R+\Delta R$, with $\Delta M$ and $\Delta R$ the 1-$\sigma$ error in mass and radius). This compact planet has at least 60\% earth-like nucleus by mass and between 0-40\% bulk water+ices content.}
   \label{ternary}
\end{center}
\end{figure}

It is not possible to show all the possible compositions for a low-mass planet using a mass-radius diagram, for this we resort to a ternary diagram (Fig.~\ref{ternary}). We computed various theoretical internal structures using the internal structure model described in \citet{2013ApJ...775...10V}, suitable for rocky and gaseous planets. Ternary diagrams relate the composition in terms of Earth-like nucleus fraction, water+ices fraction, and H-He fraction to total mass, to the radius for a specific planetary mass. Each vertex corresponds to 100\%, and the opposite side to 0\% of a particular component, by mass. The color bar shows the radius in terms of Earth radii, and the gray lines are the isoradius curves labeled in terms of Earth radii.  The possible compositions for HD\,97658\,b are in the low right corner of Fig.~\ref{ternary} (shaded in black) and correspond to a bulk composition of H-He of less than 2\%, water+ices 0-40\%, and rocks in excess of 60\%. The maximum H-He is obtained for a rock fraction of 92-95\%, and the rest water+ices.  If the planet had no water, the amount of H-He would be less than $8\times10^{-3}$ by mass. If it had no H-He, the amount of water+ices is 15-40\% by mass (the range reflects the one sigma errors in mass and radius). In comparison, GJ\,1214\,b admits at most 7\% of H-He and it could be made out of 100\% water+ices (although very unlikely). Furthermore, GJ\,1214\,b permits a very large range in rocky component (0-97\%), while HD\,97658\,b is more compact and requires a large proportion of rocks (60-99\%), thus requiring a formation mechanism that captures enough solid material.  Unfortunately, as discussed in \citet{2013ApJ...775...10V}, internal structure models do not constrain the mixing ratio between H-He and water+ices in the atmosphere. Thus, they are limited in guiding spectroscopic studies, which we discuss in the next section.  

\subsubsection{Constraints on Atmospheric Composition}

Transmission spectra of super-Earths can provide constraints on the atmospheric composition of the day-night terminator region of the planetary atmosphere, in particular on the mean molecular weight (MMW) of the atmosphere (e.g. \citealt{2009ApJ...690.1056M}), or on the presence of clouds. The atmospheric compositions of super-Earths are a subject of active debate at present, as it is unknown if their atmospheres are H/He-rich like those of ice-giants in the solar system or if they are rich in heavier molecules such as H$_2$O, CO, or N$_2$, like those of terrestrial planets. Detailed transmission spectroscopy has been reported for only one super-Earth to date, GJ~1214\,b, for which the observations indicate a flat spectrum over a wide wavelength range of $\sim$0.5-5 $\mu$m (e.g. \citealt{2010Natur.468..669B,2011ApJ...736...12B,2011ApJ...731L..40D,2012A&A...538A..46D}). The featureless spectrum of GJ~1214\,b suggests the likely presence of thick clouds in the atmosphere (\citealt{2012ApJ...756..176H,2012ApJ...745....3M,2013ApJ...778..153B,2013ApJ...775...33M,2014Natur.505...69K}) which might be obscuring the spectral features of volatile species under the cloud cover, due to which its atmospheric composition is still unknown. 

In this section, despite the only 2-$\sigma$ significance of the planetary radius discrepancy between \textit{MOST} and \textit{Spitzer} data, and the possibility of instrumental \textit{MOST} systematics not fully corrected (see Sect.5.2), we attempted to interpret the origin of the discrepancy based on atmospheric modeling of HD\,97658\,b. We also assumed here that the stellar variability of HD\,97658 is weak enough to infer atmospheric constraints directly from the transit-depth measurements. With a projected rotational velocity of 0.5 $\pm$ 0.5 km s$^{-1}$, a magnetic cycle at least 6-year long and a stellar rotation of 38.5 $\pm$ 1 d \citep{2011arXiv1109.2549H}, HD\,97658 is most probably a very quiet star. Furthermore, occulted starspots leave a clear structure in the transit light curve, which we do not observe, either with \textit{MOST} or with \textit{Spitzer}. 

The transit-depth measurements in the two bandpasses, the {\it MOST} bandpass in the visible centered at 0.525 $\mu$m and the {\it Spitzer} IRAC bandpass in the infrared at 4.5 $\mu$m, place complementary constraints on the atmospheric composition. The {\it MOST} bandpass is diagnostic of scattering phenomena as well as absorption due to a variety of chemical species, including Na, K, and particulates. On the other hand, the {\it Spitzer} 4.5 $\mu$m bandpass constrains molecular absorption, particularly due to H$_2$O, CO and CO$_2$. We used model transmission spectra of HD\,97658\,b with a wide range of compositions to interpret the observations. We considered two classes of model atmospheres: (a) cloud-free atmospheres with varied gas compositions \citep{2009ApJ...707...24M}, and (b) atmospheres with clouds/hazes of varied particulate compositions (\citealt{2012ApJ...756..176H}; see also \citealt{1999ApJ...512..843B,2001RvMP...73..719B,2007ApJS..168..140S}). These two sets of model atmospheres are known to predict slightly different amplitude variations in the 3-5 $\mu$m range for unidentified reasons (see \citealt{2012ApJ...756..176H} for a discussion about this), but this does not qualitatively impact our conclusions. 
 
\begin{figure}[!h]
\begin{center}
 \includegraphics[scale=0.6,angle=0]{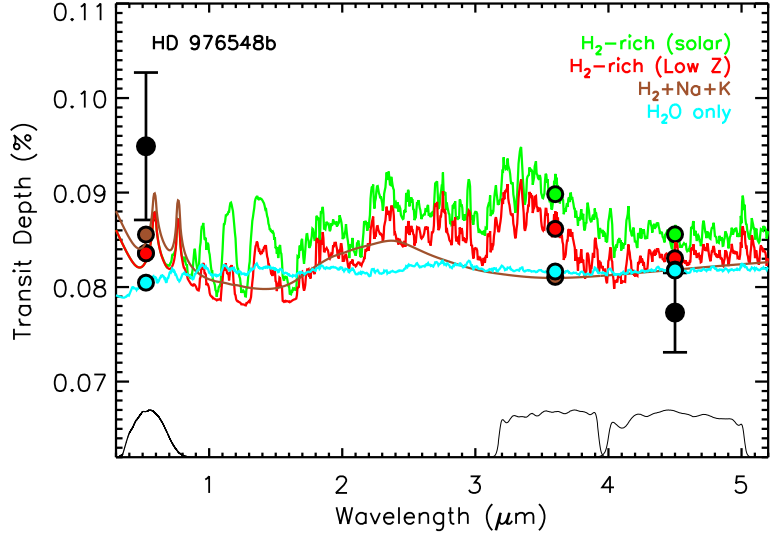} 
 \caption{Model transit spectra of super-Earth HD\,97658\,b with cloud-free atmospheres. The black circles with error bars show the two observed photometric transit depths in the {\it MOST} and {\it Spitzer} bandpasses centered at 0.525 $\mu$m and 4.5 $\mu$m, respectively. The solid black curves at the bottom show the corresponding bandpasses. The {\it Spitzer} 3.6 $\mu$m bandpass is also shown for comparison. The solid curves show four model spectra of cloud-free atmospheres with different chemical compositions, computed using the modeling approach of Madhusudhan \& Seager (2009). The green model assumes a H-rich solar composition atmosphere in thermochemical equilibrium. The green, red, and brown models are all H-rich but with different amounts of C and O. While the green model has a solar composition of C and O in chemical equilibrium, the red model is depleted in C and O by a factor of 20 relative to solar abundances and the brown model has no C and O based molecules. The blue model is a water-world atmosphere with 100\% H$_2$O. The colored filled circles are the models binned in the bandpasses. As discussed in the text, the data are inconsistent (at the 2-$\sigma$ level) with both a solar composition atmosphere as well as the H$_2$O-rich scenario. A cloud-free H-rich atmosphere depleted in C and O can explain the data marginally, at the $\sim$1-$\sigma$ uncertainties.}
   \label{atmo}
\end{center}
\end{figure}

The observations, if interpreted in the context of atmospheric modeling, can nominally be explained by a cloud-free atmosphere but only with a metal-poor composition. Figure~\ref{atmo} shows model atmospheric spectra of HD\,97658\,b with varied compositions, based on the modeling approach of \citet{2009ApJ...707...24M}. We considered models with H$_2$-rich atmospheres as well as models with high MMW, e.g. H$_2$O-rich atmospheres. We found that a cloud-free solar composition H$_2$-rich atmosphere in the planet is inconsistent with the data (green curve on Fig.~\ref{atmo}, which is entirely overlapped by the red, H$_2$-rich low metallicity model in the optical wavelength range). Such an atmosphere predicts a significantly higher 4.5 $\mu$m transit depth than observed, by over 2-$\sigma$, due to molecular absorption by H$_2$O, CO and CO$_2$ in that bandpass. On the other hand, a cloud-free high MMW atmosphere, e.g. H$_2$O-rich, is also not favored by the data, as such a model predicts significantly lower transit depth in the visible MOST bandpass than observed, by 2-$\sigma$. We found that a cloud-free H$_2$-rich atmosphere with sub-solar C and O abundances, by a factor of 20 below solar or lower, is able to match both the data points at the $\sim$1-sigma uncertainties. The low C and O abundances in such a model lower the H$_2$O, CO and CO$_2$ absorption in order to match the low 4.5 $\mu$m transit depth, while the ambient H$_2$-rich composition contributes Rayleigh scattering in the visible wavelengths which together with Na and K absorption matches the high transit depth in the {\it MOST} bandpass. While this model fits the data reasonably well, the peculiar composition with low C and O abundances poses interesting theoretical questions for future investigation. 
 
\begin{figure}[!h]
\begin{center}
 \includegraphics[scale=0.6,angle=0]{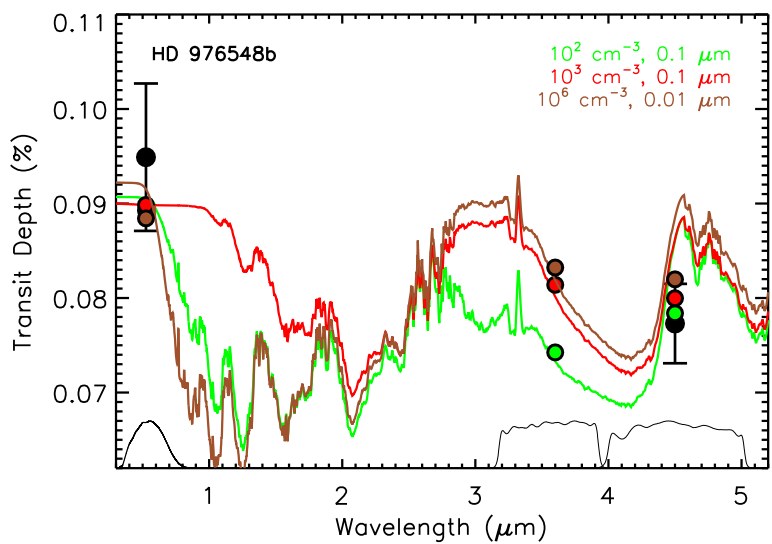} 
 \caption{Model transit spectra of super-Earth HD 97658b with atmospheres consisting of tholin haze particles. The solid curves show spectra of three model atmospheres with tholin hazes of different particle sizes and densities, based on the models of Howe \& Burrows (2012). All the models contain monodispersed tholin hazes placed in a H-rich solar-abundance atmosphere uniformly in the $10^{-4} - 10^{-6}$ bar pressure range. The particle sizes and densities for the different models are shown in the legend. All the three haze models fit the data very well within the 1-$\sigma$ uncertainties.}
   \label{atmo2}
\end{center}
\end{figure}

The observations, again if interpreted in the context of atmospheric modeling, are more readily explained by an atmosphere with significant Mie scattering due to hazes. We considered atmospheres with a wide range of haze compositions and particulate sizes based on the models of \citet{2012ApJ...756..176H}. We found that the observations can be explained by the presence of tholin hazes in a solar-abundance H$_2$-rich atmosphere, as shown in Fig.~\ref{atmo2}. The models include monodispersed tholin hazes placed uniformly in the upper atmosphere, in the $10^{-4} - 10^{-6}$ bar pressure range, in an otherwise solar-abundance atmosphere. The models shown in Fig.~\ref{atmo2} range in particle sizes between 0.01-0.1 $\mu$m and particle densities of $10^2 - 10^6$ cm$^{-3}$. All the models show a steep rise in the transmission spectrum in the visible wavelengths due to Mie scattering due to the hazes thereby explaining the high observed transit depth in the 0.525 $\mu$m {\it MOST} bandpass. Of the models shown in Fig.~\ref{atmo2}, the best fitting model contains 0.1 $\mu$m particles with a number density of $100$ cm$^{-3}$, but is only a marginally better fit relative to the two other models. 

To conclude this discussion, we stress that future observations at higher resolution and higher precision are definitely needed to confirm the atmospheric modeling interpretation as presented here, and to further constrain the atmospheric composition of HD\,97658\,b. 

\subsection{Improving the knowledge of the host star} 
An accurate knowledge of the host star is essential to derive accurate planet parameters, as illustrated here by the importance of modeling properly the stellar luminosity and radius. Age is also a very important datum, since it is expected to be equal to the age of the planet. For low-mass stars such as HD\,97658, the stellar age is essentially unconstrained by the comparison with evolutionary tracks. Based on log $R'_{\rm HK}$ and $P_{\rm rot}$ calibrations, the stellar age is around 6.0 $\pm$ 1.0 Gyr \citep{2011arXiv1109.2549H}. However, the physics behind these two empirical methods is not fully understood, and their calibrations may also present large unknown systematics that are not reflected in the 1.0 Gyr error announced \citep{2010ARA&A..48..581S}. The $UVW$ space velocities, as measured from \textit{Hipparcos} \citep{2007A&A...474..653V}, are small. This indicates that HD\,97658 has a thin disk kinematics, from which no useful constraints on its age can be obtained. Asteroseismology, the study of the oscillations of stars, can improve this situation (for the general principles of asteroseismology and some of the most recent achievements, see \citealt{2010aste.book.....A}). In solar-type (roughly mid-F to mid-K types) stars, oscillations correspond to acoustic waves (also called p-modes) that depend on the radially varying density and internal speed of sound in the star. The power pulsation spectra of solar-type stars exhibit characteristic structures, with regular spacings between the peaks. Among them, two are of utmost importance: the mean \textit{large separation} $\Delta\nu$ ($\equiv \langle\nu_{n+1,l}-\nu_{n,l}\rangle$, where $\nu$ is the frequency, $n$ the radial order, and $l$ the angular degree of the oscillation mode) and the \textit{small separations} $\delta\nu$ ($\equiv \nu_{n,l}-\nu_{n-1,l+2}$). These quantities are the first quantities that can be measured in an observed pulsation spectrum, even if the quality of the data is insufficient to extract individual p-modes frequencies \citep{2006MNRAS.369.1491R,2009A&A...506..435R,2009A&A...508..877M}. It can be shown that the large separation is approximately the inverse acoustic diameter of the star which means, from homology arguments, that it scales approximately as the square root of the mean density of the star: $\Delta\nu \propto \sqrt{\rho_*}$. For their part, the small separations $\delta\nu$ provide a measure of conditions in the core of the star and hence, the stellar age.

\begin{figure}[!h]
\begin{center}
\includegraphics[scale=0.5,angle=0]{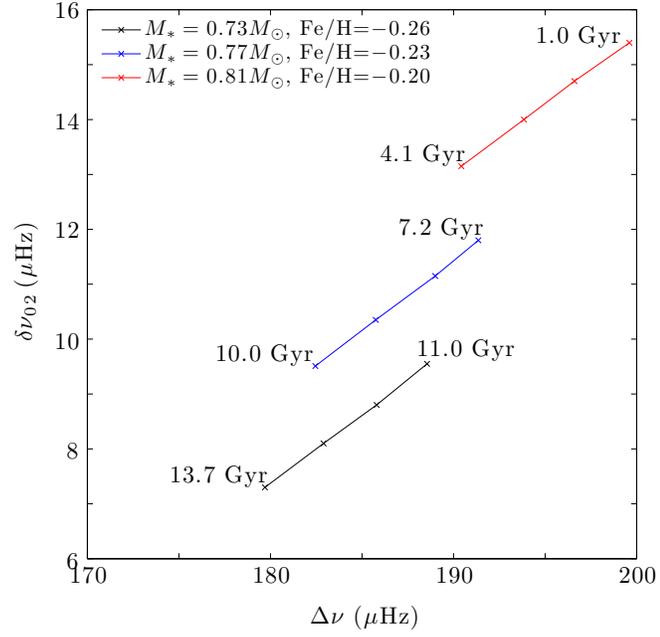}\\ 
 \caption{C-D diagram (large separations $\Delta\nu$ \textit{vs} small separations $\delta\nu_{02}$) for stellar models with various masses and metallicities, but that are consistent with the observational constraints ($T_{\rm eff}, L_*$, [Fe/H]) on HD\,97658.}
   \label{fig2}
\end{center}
\end{figure}

To illustrate the potential of asteroseismology to improve the knowledge of the host star, in particular by constraining its age, we computed the seismic properties of 12 selected stellar models with the Liege adiabatic pulsation code OSC \citep{2008Ap&SS.316..149S}. These models have various masses, metallicities, and ages, and all simultaneously respect the observational triplet ($T_{\rm eff}, L_*$, [Fe/H]) within the associated 1-$\sigma$ uncertainties. Their properties are presented in Table \ref{tab3}. We plotted in Fig.~\ref{fig2} the $\Delta\nu - \delta\nu_{02}$ diagram, called the C-D diagram in the asteroseismic jargon \citep{1984srps.conf...11C}, for these 12 stellar models\footnote{Here $\Delta\nu = \nu_{n+1,0}-\nu_{n,0}$ and $\delta\nu_{02} = \nu_{n,0}-\nu_{n-1,2}$. They are given for the corresponding frequency at maximum power, around which the pulsation spectrum is expected to be observed.}. Figure~\ref{fig2} shows that an accuracy of $\sim$1 $\mu$Hz on the large and small separations measurements will help to get an improved knowledge on the stellar mass and age. The main unknown for asteroseismic observations concerns the amplitudes of the oscillations, which are not predictable by the current linear theory of stellar oscillations. The amplitudes are expected to be weak in K-type stars \citep{1995A&A...293...87K}, but HD\,97658 is a quite bright star. We estimated, from the measured CoRoT performances on similar type and magnitude stars (see the example of the K0 dwarf $V=7.84$ HD 46375, \citealt{2010A&A...524A..47G}) that a space-borne 0.3-m mission like CHEOPS \citep{2013EPJWC..4703005B} would be able to obtain such a precision from $\sim$2-3 months of nearly continuous observations. This may be a suggestion for the complementary CHEOPS open program. The PLATO mission, if HD\,97658 falls in the observed fields, will definitely be able to measure the large and small separations with the required accuracy. PLATO should even be able to accurately measure the individual oscillation frequencies, from which a full asteroseismic analysis can be carried out, obtaining not only very accurate global parameters (stellar mass, radius, age) but also constraints on the internal physics of the star. This is in turn important to calibrate and improve the evolutionary tracks computed from stellar evolution codes. 

\section{Conclusion and Prospects}
\label{cc}
HD\,97658\,b is a key transiting super-Earth. The current data from \textit{Spitzer} and \textit{MOST} photometry, and also Keck-Hires RVs, analyzed with our MCMC code, indicate a planet with an intermediate average density ($\rho_P=3.90^{+0.70}_{-0.61}$ g cm$^{-3}$). Investigating the possible internal compositions of HD\,97658\,b, our results indicate a large rocky component by at least 60\% by mass, an amount of 0-40\% of water+ices, and very little H-He components, at most 2\% by mass. If interpreted as constraints on the atmospheric composition, the transit depths in both the \textit{Spitzer} and \textit{MOST} bandpasses together favor either a H$_2$-rich atmosphere with hazes or a cloud-free atmosphere with a low metallicity. Six more transit observations are planned in the course of 2014 with the \textit{Spitzer} IRAC camera, this time with 3.6 $\mu$m channel (PI: D. Dragomir).
%, and a non-negligible envelope made of water+ices and/or H-He. 

HD\,97658\,b will be a target of the coming space missions TESS and CHEOPS, in particular to accurately measure the planet radius at visible wavelengths. By the time TESS and CHEOPS are launched (circa 2017), the GAIA mission\footnote{http://sci.esa.int/gaia} (launched in December 2013) is hoped to have provided extremely accurate parallax measurements, within $\sim$ 10 $\mu$as for the bright nearby stars such as HD\,97658. This will improve the knowledge on the distance of the star by 2 orders of magnitude compared to the current \textit{Hipparcos} parallax. From there, the stellar radius will be known with an accuracy of 1\% (assuming a $\pm$ 50 K uncertainty on $T_{\rm eff}$).  Knowing the stellar radius to 1\% is also what we can expect for present and near-future long baseline interferometers \citep{2013ApJ...771...40B}. This is the ultimate precision we can achieve on the planet radius, providing that the number of observed transits is sufficient to measure the transit depth $dF \propto (R_P/R_*)^2$ with an accuracy much below 1\%. ESA has now officially confirmed the selection of PLATO as the next M3 mission, to be launched around 2024. If HD\,97658 falls in the observed fields, PLATO would not only measure the transit depth with unprecedented accuracy, but would also be able to detect oscillation frequencies of the host star that provide through asteroseismology accurate stellar mass and, very importantly, age.

Orbiting a bright K1-type star ($V=7.7$, $K=5.7$), HD\,97658\,b is a target of choice for atmospheric characterization. HST time is allocated to take in the coming months transmission spectroscopy of HD\,97658\,b with the WFC3 camera in IR (PI: H. Knutson). The aim is to distinguish a large scale-height, H-dominated atmosphere, from a compact, water steam atmosphere, which will confirm/refute the constraints on the atmospheric composition derived in this work. By comparing these measurements to the few targets for which such transmission spectroscopy is currently possible (55 Cnc e and GJ 1214\,b), this will provide the first measure of the diversity of super-Earth atmospheres prior the era of the European Extremely Large Telescope (E-ELT) and the James Webb Space Telescope (JWST).
% \citet{2006SSRv..123..485G}  Coming observations by HST (transmission spectrum).
%discuter future obs : Cheops, TESS, . GAIA. 
%Atmosphere ? Detection of the occultation (secondary eclipse), when the planet is hidden by the host star. 

\begin{acknowledgements}
V. Van Grootel warmly thanks A. Grotsch-Noels and M.-A. Dupret for fruitful discussions on stellar modeling. The authors also thank H. Knutson for various discussions on this planet. This work is based in part on observations made with the {\it Spitzer Space Telescope}, which is operated by the Jet Propulsion Laboratory, California Institute of Technology under a contract with NASA. Support for this work was provided by NASA. M. Gillon is  Research Associate at the Belgian Fonds de la Recherche Scientifique (FNRS). 
\end{acknowledgements}

{\it Facilities:} \facility{Keck:I (HIRES)}, \facility{\textit{MOST}}, \facility{\textit{Spitzer}}

\bibliographystyle{aastex}
\bibliography{Biblio/references}

\clearpage

  \begin{sidewaystable}[!ht]
 \begin{center}
 {\tiny \begin{tabular}{c c c c c c c c c c c}\hline
\hline
BJD\_UTC&$Flux$& Error & dX& dY & FWHM & FWHM\_X&FWHM\_Y&sky&airmass&Exposure time\\
d&---&---&pix&pix&pix&pix&pix&$e^{-}$&---&s\\
 \hline 
   6523.0421407212862     &   1.0004247181036807    &   6.56049340879199779E-004  & 15.998328125000002     &   16.065718749999998    &    1.1364000000000001     &   1.2856813328125001        & 1.3256666562500001   &     3.6243129999999999      & 1 &  8.00000000000000017E-002 \\
     6523.0422367210858   &     1.0000111640826093     &  6.42787543238822440E-004 &  16.010062499999997    &    16.060171874999998    &    1.1312000000000000      &  1.2750128125000000        & 1.3148729515625006    &    2.4829750000000002  &     1 &   8.00000000000000017E-002 \\
 6523.0423342212462 &      0.99991532572716002  &     5.79726271470056421E-004 &  16.009390625000002   &     16.055859374999997   &     1.1294000000000000     &   1.2703478468750000        &1.3100962265625005    &    1.6571240000000000     &  1 &   8.00000000000000017E-002 \\
   6523.0424312212863    &    1.0012275914818374  &     6.76242623735089695E-004   &16.006359374999999     &   16.065999999999999  &      1.1361000000000001   &     1.2820394375000006        & 1.3274429203125000    &    4.4898300000000004  &     1 &   8.00000000000000017E-002 \\
   6523.0425283214436  &      0.99966334756166497   &    6.49723785530624651E-004  & 15.985859374999997  &      16.071609374999994  &      1.1443000000000001   &     1.3029140687500000        & 1.3381304734374999  &       3.0959590000000001   &    1 &   8.00000000000000017E-002 \\
   6523.0426265214928     &   1.0002273840403539     &  7.43672238581453439E-004  & 15.979874999999998   &     16.060750000000002   &     1.1417999999999999  &      1.3023067859375002        & 1.3236822687499994    &    5.3868869999999998   &    1 &   8.00000000000000017E-002 \\
   6523.0427228211784    &    1.0005791404475077    &   7.23806840153553297E-004  & 15.987656250000004  &      16.052609374999992    &    1.1341000000000001   &     1.2907193265625003        & 1.3096692703124997   &     4.1206699999999996  &     1 &   8.00000000000000017E-002 \\
   6523.0428201211043    &    1.0010497355426586     &  5.82884097738351685E-004  & 15.984656249999995  &      16.038562500000001   &     1.1333000000000000   &     1.2924486984374999        & 1.2930500937500000  &      4.8478009999999996 &      1 &  8.00000000000000017E-002 \\
   6523.0429173213788   &    0.99945788748303044  &     6.51979662206761119E-004 &   15.990640624999997   &     16.075765625000003     &   1.1457999999999999    &    1.3031341859375003        & 1.3458211124999993  &      4.4744950000000001   &    1 &  8.00000000000000017E-002 \\
   6523.0430144210704   &     1.0002340513003241  &     6.45661570025192044E-004  & 16.011250000000000     &   16.070406249999998   &     1.1361000000000001   &     1.2846436531250001        &1.3306431562499996   &     1.6783100000000000    &   1 &   8.00000000000000017E-002 \\
 \hline
 \end{tabular}}
 \end{center}
 \caption{\textit{Spitzer} photometric time series of HD\,97658, as used by our MCMC algorithm.}\label{rawdata}
 \end{sidewaystable}

 \begin{table*}[!ht]
 \begin{center}
 {\small \begin{tabular}{l c c c}\hline
 \hline
 Parameter&Symbol&Value&Unit\\
 \hline 
% \hline
Jump parameters&&&\\
\hline
\multicolumn{4}{l}{\textit{Jump parameter, uniform prior}}\\ 
 Transit depth, \textit{Spitzer}&$dF$&$773 \pm 42$&ppm\\
 Transit width&$W$&$0.1187 \pm 0.0012$&days\\
 Mid-transit time-2450000&$T_0$&$6523.12540^{+0.00060}_{-0.00056}$&BJD\_TDB\\
 Impact parameter&$b'=a \cos i / R_*$&$0.35^{+0.13}_{-0.21}$&$R_*$\\
 Orbital period&$P$&$9.4903^{+0.0016}_{-0.0015}$&days\\
 &$\sqrt{e} \cos \omega$&$0.05^{+0.18}_{-0.20}$&\\
 &$\sqrt{e} \sin \omega$&$0.18^{+0.13}_{-0.23}$&\\
 &$K_2$&$5.76^{+0.56}_{-0.58}$&\\
\multicolumn{4}{l}{\textit{Jump parameter, Gaussian prior}}\\
Stellar effective temperature&$T_{\rm eff}$&$5170\pm50$&K\\
Stellar metallicity&$\rm [Fe/H]$&$-0.23\pm0.03$&\\
Stellar luminosity&$L_*$&$0.355\pm0.018$&$L_{\odot}$\\
Stellar mass&$M_*$&$0.77\pm0.05$&$M_{\odot}$\\
 \hline
Derived stellar parameters&&\\
 \hline
Stellar radius&$R_*$&$0.741^{+0.024}_{-0.023}$&$R_{\odot}$\\
Stellar density&$\rho_*$&$1.89^{+0.23}_{-0.20}$&$\rho_{\odot}$\\
Stellar surface gravity&$\log g_*$&$4.583^{+0.047}_{-0.054}$&\\
Limb darkening coeff.&$u_1$&$0.07313^{+0.00079}_{-0.00078}$\\
Limb darkening coeff.&$u_2$&$0.1442\pm0.0013$\\
Distance&$d$&21.11 $\pm$ 0.34& pc\\
\hline
 Derived planet parameters&&\\
 \hline
Radii ratio&$R_P/R_*$&$0.02780^{+0.00075}_{-0.00077}$&\\
Planet radius (at 4.5$\mu$m)&$R_P$&$2.247^{+0.098}_{-0.095}$&$R_{\oplus}$\\
Planet mass&$M_P$&$7.55^{+0.83}_{-0.79}$&$M_{\oplus}$\\
Planet density&$\rho_P$&$3.90^{+0.70}_{-0.61}$&g cm$^{-3}$\\
Planet surface gravity&$\log g_P$&$3.166^{+0.059}_{-0.061}$&\\
Orbital inclination&$i$&$89.14^{+0.52}_{-0.36}$&deg\\
Orbital semi-major axis&$a$&$0.080^{+0.0017}_{-0.0018}$&AU\\
Orbital eccentricity &$e$&$0.078^{+0.057}_{-0.053}$&\\
Argument of the periastron&$\omega$&$71^{+65}_{-63}$&deg\\
RV orbital semi-amplitude &$K$&$2.73^{+0.26}_{-0.27}$&m/s\\
Planet equilibrium temperature& $T_{\rm eq}$&$757^{+12}_{-13}$&K\\
 \hline
 \end{tabular}}
 \end{center}
 \caption{Median and 1-$\sigma$ limits of the posterior distributions derived for HD\,97658 and its planet from our MCMC analysis of \textit{Spitzer} photometry and Keck-HIRES RVs.}\label{tab1}
 \end{table*} 
 
  \begin{table}[!ht]
 \begin{center}
 {\small \begin{tabular}{c c c c c c c c}\hline
\hline
Model&$M_*$& Age &$T_{\rm eff}$&$L_*$&[Fe/H]&$\Delta\nu$&$\delta\nu_{02}$\\
&($M_{\odot}$)& (Gyr)&(K)&($L_{\odot}$)&&($\mu$Hz)&($\mu$Hz)\\
 \hline 
 1&0.73&11.0&5144&0.337&$-$0.26&188.5&9.5\\
 2&0.73&11.9&5165&0.348&$-$0.26&185.8&8.8\\
 3&0.73&12.8&5187&0.362&$-$0.26&182.9&8.1\\
 4&0.73&13.7&5210&0.377&$-$0.26&179.7&7.3\\
 5&0.77&7.2&5130&0.338&$-$0.23&191.4&11.8\\
 6&0.77&8.0&5144&0.346&$-$0.23&189.0&11.2\\
 7&0.77&9.0&5170&0.361&$-$0.23&185.8&10.3\\
 8&0.77&10.0&5195&0.377&$-$0.23&182.4&9.5\\
 9&0.81&1.1&5152&0.337&$-$0.20&199.6&15.4\\
 10&0.81&2.0&5171&0.346&$-$0.20&196.6&14.7\\
 11&0.81&2.9&5192&0.358&$-$0.20&193.8&14.0\\
 12&0.81&4.1&5220&0.375&$-$0.20&190.4&13.2\\
 \hline
 \end{tabular}}
 \end{center}
 \caption{Mass, age, effective temperature, luminosity, metallicity, large separations, and small separations of 12 stellar models consistent with the observational constraints ($T_{\rm eff}, L_*$, [Fe/H]).}\label{tab3}
 \end{table} 

\clearpage

\clearpage

\clearpage

\clearpage

\clearpage

\clearpage

\clearpage 

\end{document}